\documentclass{elsart}
\usepackage{epsfig}
\usepackage{color}
\usepackage{amsmath}
\usepackage{amsfonts}
\usepackage{graphicx,slashbox}
\usepackage{ulem}
\usepackage{cite}

\definecolor{green2}{rgb}{0,0.5,0}

\begin{document}

\begin{frontmatter}

\title{The Generalized SIC-OEP formalism and the Generalized SIC-Slater approximation (stationary and time-dependent cases).}
\author{J. Messud$^{a,b}$, P.~M.~Dinh$^{a,b}$},
\author{P.-G.~Reinhard\corauthref{cor}$^c$},
\author{E.~Suraud$^{a,b,c}$} 

\corauth[cor]{Corresponding author\\{\it Email-address}~:
  Paul-Gerhard.Reinhard@theorie2.physik.uni-erlangen.de}
\address{$^a$Universit\'e de Toulouse; UPS; \\Laboratoire de Physique
  Th\'eorique (IRSAMC); F-31062 Toulouse, France}
\address{$^b$ CNRS; LPT (IRSAMC); F-31062 Toulouse, France}
\address{$^c$Institut f{\"u}r Theoretische Physik, Universit{\"a}t
  Erlangen, D-91058 Erlangen, Germany}

\begin{abstract}
We present a generalized formulation of the Optimized
Effective Potential (OEP) approach to the Self Interaction
Correction (SIC) problem in Time Dependent (TD) Density Functional
Theory (DFT). The formulation relies on the introduction of a double
set of single electron orbitals. It allows the derivation of a
generalized Slater approximation to the full OEP
formulation, which extends the domain of validity of the standard
Slater approximation. We discuss both formal aspects and practical
applications of the new formalism and give illustrations in cluster
and molecules.  The new formalism provides a valuable ansatz to more
elaborate (and computationally very demanding) full TD OEP and full
TD SIC calculations especially in the linear domain.
\end{abstract}

\end{frontmatter}

%\tableofcontents

\section{Introduction}
\label{sec:intro}

Density Functional Theory (DFT) has evolved to be one of the
most powerful theoretical frameworks for the description of complex
chemical and physical systems.
%~\cite{Hoh64,Par89,Dre90,Koh99r}. 
Enormous progress has been made since the seminal works of the sixties
by Kohn et al.~\cite{Hoh64,Koh65}. DFT is now routinely used,
especially in systems with a large number of electrons
\cite{Par89,Dre90,Koh99r}. Nevertheless, there remain still several
open questions in detail which are in the focus of actual research
\cite{last_dft_meeting}.  The
extension to Time-Dependent DFT (TDDFT) has been formally established
more recently \cite{Run84,Gro90,Mar04}. It is still a developing
field, at the side of both formal and practical aspects \cite{Mar06}.
Already at the present stage, TDDFT has become one of the few well
founded theories for describing the dynamics of complex systems. 
This is especially true for non equilibrium situations such as 
clusters and molecules under the influence of intense laser
fields \cite{Rei03a}.

DFT simplifies the involved problem of many-electron correlations in
terms of an effective (Hartree-like) one-body description. This is
achieved by introducing exchange and correlation effects in an energy
functional expressed in terms of the local density of the
electrons. The simplest strategy along that line is provided by the
Local Density Approximation (LDA) which has proven in many
calculations to provide a simple and reliable description of structure
and low-amplitude excitations (optical response, one-photon processes)
\cite{Koh99r}. The analogue in the time-dependent case is the
Adiabatic Local Density Approximation (ALDA) which has also been used
with great success in dynamical processes involving huge energy
deposits and/or large ionization of irradiated clusters and molecules
\cite{Rei03a}.

However, the LDA is plagued by a self-interaction error because the
direct Coulomb term and the LDA exchange-correlation potential involve
the total density including the particle on which the mean-field acts
in the Kohn-Sham equations. While in a full Hartree-Fock treatment, the
exchange term exactly cancels the self interaction of the direct term,
the approximate treatment of the exchange term in LDA destroys this
cancellation. As a consequence, a self-interaction remains and one of
the defects is that LDA produces a wrong Coulomb asymptotics
\cite{Per81,Dre90}. The self-interaction thus spoils single-particle
properties in particular the Ionization Potential (IP) in finite
systems or the band gap in solids \cite{Hyb86a,Nie00aR}. It is also
well known that LDA fails in describing the polarizability in chain
molecules \cite{Gis99a,Kue04a}. In the dynamical case, the
self-interaction error also spoils the description of ionization
dynamics, especially close to threshold where IP effects dominate.

There exist ways to correct the self-interaction error while trying to
keep the simplicity of the method.  An early attempt along that line was proposed
by Fermi and Amaldi \cite{Fer34}. The standard way to introduce a Self
Interaction Correction (SIC) is based on the more recent proposal of
Perdew \cite{Per79a,Per81}. Such SIC has been explored since then at
various levels of refinement for structure calculations in atomic,
molecular, cluster and solid state physics, see
e.g. \cite{Ped84,Goe97,Polo,Vyd04}. The SIC scheme, however, leads to
an orbital dependent mean-field which causes several formal and
technical difficulties. A way out is provided by using optimized
effective potentials (OEP) techniques as introduced in
\cite{Sha53,Tal76}, see~\cite{Kue07} for a recent review. However,
applying OEP to SIC leads to a very involved formalism usually treated
with further approximations, as e.g. the Krieger-Li-Iafrate (KLI)
approach~\cite{Kri92a,Kri92b}. But these approximations can severely
perturb some crucial physical features of SIC, particularly the trend
to produce localized single-particle orbitals \cite{Kue07}. It is thus
a key issue to refine such approximate schemes to SIC in order to
preserve, as much as possible, original SIC properties, at a lower
cost than full OEP. 
It should be noted that there is nevertheless a further advantage of OEP.  It
optimizes one local mean-field Hamiltonian for the system.  This
allows to evaluate unambiguously unoccupied states of the system
which, in turn, can have important applications in dynamical
processes.

Time-dependent situations call for a time-dependent SIC (TDSIC).
Applications of TDSIC have, up to now, mostly been performed in
approximate manners, e.g., the {linearized} treatment of \cite{Pac92},
averaged-density SIC \cite{Leg02} based on a generalization of the
Amaldi picture \cite{Fer34}, or the various versions of time-dependent
OEP-KLI \cite{Ull95a,Ton97,Ton01}. Only recently, a manageable and
exact propagation scheme for TDSIC has been formulated
\cite{Mes08a,mes09} which is applicable in all dynamical ranges.  The
key to success is to employ two complementing sets of occupied single
particle wave functions, one for the mean-field propagation and the other
one establishing the necessary localization of the wave functions.
The double-set technique has also proven to be extremely useful to
formulate efficient approximations to OEP. This was demonstrated for
the stationary case in \cite{Mes08b,mes09-3} and later on extended to
the time domain \cite{Mes09-4}.  The aim of the present paper is to
present and discuss in more detail the local approximations to time
dependent OEP (TDOEP) based on the double-set technique.

The paper is organized as follows. 
In section \ref{sec:TDSIC}, we briefly review the double-set technique
for SIC and TDSIC. 
In section \ref{sec:TDOEP}, we introduce the (TD)OEP equations in the
light of the double-set representation and develop from that what we
call the generalized Slater approximation to OEP.
In section \ref{sec:numres}, we present results for a variety of test
cases, static as well as dynamic ones, and compare with results from
the lower approximations LDA and ADSIC, and from full TDSIC as
a benchmark.
Conclusions are summarized in section \ref{sec:concl}.

\section{Brief review of the double-set technique for SIC}
\label{sec:TDSIC}

In this section, we give a brief outline of SIC and TDSIC in the
double-set formulation as introduced in \cite{Mes08a,mes09}. We start
from the static case which helps to motivate the double-set technique
and proceed to the dynamical case where the use of two sets of
wave functions is compulsory. The brief review of (TD)SIC should serve
as a starting point for the derivation of improved approximations to
OEP (both in stationary and time-dependent cases).

\subsection{Stationary case}

The starting point is the SIC energy functional for electrons~:
\begin{eqnarray}
  E_\mathrm{SIC}[\{\psi_\alpha\}] =
  \sum_\alpha \left(\psi_\alpha \big|\frac{\mathbf{p}^2}{2m}\big|\psi_\alpha \right)
  \!+\!
  E_\mathrm{ext}[\rho]
  \!+\!
  E_\mathrm{LDA}[\rho]
  \!-\!
  \sum_\alpha E_\mathrm{LDA}[|\psi_{\alpha}|^2] \ ,
\label{eq:fsicen}
\end{eqnarray}
whereby all sums run over occupied states only. Note that we omit the
space-time dependencies, i.e. $\psi_\alpha=\psi_\alpha(\mathbf{r},t)$,
when it is 
not misleading. $\rho$ stands for the total electronic density
$\rho=\sum_\alpha \rho_\alpha = \sum_\alpha|\psi_\alpha|^2$. 
The first term in $E_\mathrm{SIC}$ is the {non-interacting} kinetic
energy; $E_\mathrm{ext}[\rho]=\int d\mathbf{r}\rho v_{\rm ext}$
collects all external one-body fields where $v_{\rm ext}$ stands for
the interaction with the ionic background and any other local possibly
time-dependent external field; and $E_\mathrm{LDA}[\rho]$ is a
standard LDA energy-density functional including the direct term of
the electron-electron Coulomb interaction.  The last term corresponds
to the SIC.  We mention in passing that the SIC, and with it all our
following development, does also apply to more general functionals as,
e.g., the Generalized Gradient Approximation (GGA) \cite{Per96}.
A basic assumption beyond this SIC functional is that the employed
single-particle wave functions are ortho-normalized
\begin{equation}
  (\psi_\alpha|\psi_\beta)
  =
  \delta_{\alpha\beta}
  \quad.
\label{eq:orthnorm}
\end{equation}

The stationary SIC equations are obtained by variation of the SIC
energy (\ref{eq:fsicen}) with the additional constraint on
ortho-normalization (\ref{eq:orthnorm}) which is taken into account
through the Lagrange multipliers $\lambda_{\alpha\beta}$.  This leads
to the following mean-field equations~\cite{Ped84,Mes08a,mes09}
\begin{subequations}
\label{eq:onesetSIC}
\begin{eqnarray}
  \hat{h}_\mathrm{SIC} |\psi_{\alpha})
  &=&
  \sum_{\beta} \lambda_{\alpha\beta}  |\psi_{\alpha})
  \quad,
\label{eq:static-nondiaq} 
\\
  \hat{h}_\mathrm{SIC}
  &=&
  \hat{h}_\mathrm{LDA}
  -
   \hat{U}_{\rm SIC}
  \quad,
\label{eq:hsic}
\\
  \hat{h}_\mathrm{LDA}
  &=&
  \frac{\hat{p}^2}{2m}
  +
  U_{\rm LDA} [\rho]
  \quad,
\label{eq:SICmfham}
\\
  U_{\rm LDA} [\rho]
  &=&
  \frac{\delta E_\mathrm{LDA}[\rho]}
       {\delta \rho(\mathbf{r})}
  \quad,
\label{eq:SICmf}
\\
   \hat{U}_{\rm SIC}
   &=&
  \sum_\alpha U_{\rm LDA} [ |\psi_\alpha|^2]|\psi_\alpha)(\psi_\alpha|
  \quad,
\label{eq:Usic}
\end{eqnarray}
combined with what we call the ``symmetry condition''
\begin{eqnarray}
  0
  =
  (\psi_\beta|U_{\rm LDA} [ |\psi_\beta|^2 ]-U_{\rm LDA} [|\psi_\alpha|^2 ]|\psi_\alpha)
  \quad.
\label{eq:symcond2}
\end{eqnarray}
\end{subequations}
The first term in $\hat{h}_\mathrm{SIC}$ is the standard LDA mean
field Hamiltonian (\ref{eq:SICmfham}) and the second term stems from
the SIC in the energy (\ref{eq:fsicen}). 

Thus far, Eqs. (\ref{eq:onesetSIC}) comprise the complete
stationary SIC method. The right-hand side of
Eq. (\ref{eq:static-nondiaq}) is unconventional and inconvenient as it
does not lead explicitly to single-particle energies. The latter may be
defined a posteriori by diagonalizing the matrix of Lagrange
multipliers $\lambda_{\alpha\beta}$. 
We now introduce explicitly a second set of wave functions $\{|\varphi_i)\}$
which indeed diagonalizes the SIC Hamiltonian, 
\begin{eqnarray}
  \hat{h}_\mathrm{SIC}|\varphi_i)
  =
  \varepsilon_i |\varphi_i) \ ,
\label{eq:static-diaq}
\end{eqnarray}%
and which is related to the previous set of $\{|\psi_\alpha)\}$ by a unitary
transform amongst occupied states only~:
\begin{eqnarray}
\label{eq:unitrans}
  \psi_\alpha  =  \sum_{i} \varphi_i \, u_{i\alpha}
  \quad,\quad
  \sum_{i} u_{i\alpha} u^*_{i\beta}=\delta_{\alpha\beta}
  \quad.
\end{eqnarray}
Both sets lead to the same total density $\rho$ such that the LDA
mean-field $U_{\rm LDA} [\rho ]$ remains the same.  The new set 
$\{|\varphi_i)\}$ represents the energy diagonal states, while the old
set $\{|\psi_\alpha)\}$ remains the decisive ingredient in the
symmetry condition (\ref{eq:symcond2}).  The coefficients
$u_{i\alpha}$ of the unitary transformation (\ref{eq:unitrans}) for
given $\varphi_i$ are to be determined such that the symmetry
condition (\ref{eq:symcond2}), involving the $\psi_\alpha$, is
fulfilled.  As the $\varphi_i$ 
orbitals satisfy eigenvalue equations, they are interpreted as single
electron orbitals. The set $\psi_\alpha$ serves to minimize the SIC energy
(\ref{eq:fsicen}) and to calculate the SIC mean-field
$\hat{h}_\mathrm{SIC}$.  

This completes the double-set representation of stationary SIC.  The
double-set technique is not compulsory for the stationary case, but
enlightening. The two sets play different roles. The energy diagonal
states can easily be delocalized and are likely to spread over the
whole system, e.g., when considering the valence shell of metallic
bonds. The SIC set $\{|\psi_\alpha)\}$, on the other hand, aims to
minimize the SIC energy which is usually achieved by localization of
the associated density $|\psi_\alpha|^2$ to minimize the Coulomb
energy \cite{Ped84,Ped85,mes09}.

\subsection{Time-dependent case}

In contrast to the static case, the double-set technique is a
necessary ingredient for developing a well-defined and manageable
propagation scheme.

To derive the TDSIC equations, we start from the SIC quantum action
\begin{eqnarray}
 A_{\rm SIC}
 =
 \int_{t_0}^{t_1} \textrm dt \Big(
  E_\mathrm{SIC}[\{\psi_\alpha\}](t)
  -
 \sum_{\alpha}(\psi_\alpha(t)|\mathrm{i}\hbar \partial_t|\psi_{\alpha}(t))
 \Big)
 \quad.
\label{eq:varprinconstr}
\end{eqnarray}
The situation with the action for time-dependent variation is, in
fact, not so trivial for the derivation of DFT. There can arise
problems with causality \cite{Gro95a} and boundary conditions
\cite{Lee98} for which solutions are discussed in \cite{Lee98,Vig08}.
We are dealing here with a local and instantaneous ALDA functional
which allows to use the naive action (\ref{eq:varprinconstr}).  
Moreover, concerning the theorems derived from symmetries of the action,
it is shown in \cite{Vig08} that, as compensations occur, the
stationarity of the naive action (\ref{eq:varprinconstr}) leads to the
correct final results. We thus perform variation of this action
including once again the ortho-normality constraint with Lagrange
multipliers $\lambda_{\gamma\beta}$, i.e. we require
\begin{eqnarray}
  \delta \Big( A_{\rm SIC} - \int_{t_0}^{t_1} dt \sum_{\beta,\gamma}^{}(\psi_{\beta}(t)|\psi_{\gamma}(t))
  \lambda_{\gamma\beta}(t) \Big) = 0
  \quad.
  \nonumber
\end{eqnarray}
It is to be noted that, to derive the time-dependent OEP
  formalism, one should use the action (\ref{eq:varprinconstr}) in the
  limit $t_0 \rightarrow -\infty$.  This is necessary to recover in
  the stationary limit the stationary OEP formalism, as proved in
  \cite{Gro94}.

The steps of the variation are explained in detail elsewhere
\cite{Mes08a,mes09}.  We summarize the resulting equations.  They
again employ the two sets of occupied single-particle wave functions 
which are connected by a unitary transformation (\ref{eq:unitrans}).
The set $\{|\varphi_i(t))\}$, which was the diagonal set in the static
case, now turns out to be the ``propagating set'' obeying the
time-dependent mean-field equation
\begin{eqnarray}
  \Big( \hat{h}_\mathrm{SIC}(t) - i \hbar \partial_t \Big)
  |\varphi_i(t)) 
  = 
  0 \ ,
\label{eq:td-diag}
\end{eqnarray}
where $\hat{h}_\mathrm{SIC}$ is defined by (\ref{eq:Usic}). 
The coefficients $u_{i\alpha}$ of the unitary transform
(\ref{eq:unitrans}) for given $\varphi_i$ are to be determined such
that the ``localizing set'' $\{|\psi_\alpha(t))\}$ satisfies the
symmetry condition
\begin{eqnarray}
  u_{i\alpha}(t) \quad : \quad
  \forall t , \hspace{1mm} 
  0
  =
  (\psi_\beta(t)|U_{\rm LDA} [|\psi_\beta|^2](t)
                -U_{\rm LDA} [ |\psi_\alpha|^2 ](t)|\psi_\alpha(t))
\label{eq:td-symcond2}
\end{eqnarray}
at any time.
The solution scheme for these two coupled equations is obvious.  The
time-dependent Schr\"odinger equation (\ref{eq:td-diag}) for the
propagating set is solved for a short time step by standard
techniques, e.g., a Taylor expansion  of the formal solution
$
|\varphi_i(t'))
  =
  \exp{\left\{
    -\frac{\mathrm{i}}{\hbar}\int_t^{t'}\, \textrm d\tau\,
     \hat{h}_{\rm SIC}(\tau)\right\}}
  |\varphi_i(t))
  .
$
At each time step, the set $\psi_\alpha$ is determined by resolving
the symmetry condition (\ref{eq:symcond2}) \cite{mes09}, which is
an instantaneous equation. Then the $\psi_\alpha$ serve to
construct the new mean-field $\hat{h}_{\rm SIC}$ for the next time step.

This TDSIC propagation scheme looks formally straightforward. However,
it contains one especially numerical expensive ingredient:  The
iteration of the symmetry condition (\ref{eq:symcond2}) 
requires to invoke the time-consuming Coulomb solver in each iteration
step.  
Any acceptable approximate solutions
are thus welcome. The time-dependent generalized Slater approximation
which will be discussed below is a step into this direction.

\section{SIC-OEP and the Generalized SIC-Slater approximation}
\label{sec:TDOEP}

\subsection{Stationary formalism}
\label{sec:GS_stat}

\subsubsection{SIC-OEP in double-set representation}
\label{sec:OEPstatderv}

The ``Optimized Effective Potential'' (OEP) formalism is the tool of
choice to find the best local approximation to a non-local
Hamiltonian. In the present case in which we plan to apply OEP to the
SIC problem, we start from the total SIC energy (\ref{eq:fsicen})
formulated in terms of the (localized) $\psi_\alpha$ orbitals and we
complement this set by  the diagonal orbitals $\varphi_i$. The latter
are required from the onset to satisfy a 
local eigenvalue equation
\begin{equation}
  \Big(\hat{h}_\mathrm{LDA}-U_\mathrm{SIC}^\mathrm{(local)}(\mathbf{r})\Big) 
  \varphi_i(\mathbf{r}) 
  = 
  \epsilon_i \varphi_i(\mathbf{r})
  \quad.
\label{eq:OEP10}
\end{equation}
Locality is imposed by the fact that
$U_\mathrm{SIC}^\mathrm{(local)}(\mathbf{r})$ is a function of
$\mathbf{r}$ only. The localizing set of $\psi_\alpha$ is obtained by
the unitary transformation (\ref{eq:unitrans}) whose coefficients are
optimized to minimize the SIC energy (\ref{eq:fsicen}). It remains to
determine the space of occupied single-particle states in terms of
the $\varphi_i$. The condition (\ref{eq:OEP10}) shifts the problem to
a yet unknown optimizing local potential
$U_\mathrm{SIC}^\mathrm{(local)}$. This potential then becomes the
variational degree of freedom instead of the $\varphi_i$.
The potential $U_\mathrm{SIC}^\mathrm{(local)}(\mathbf{r})$ is thus determined
%as for the exact SIC formalism, 
by minimization of the total SIC energy (\ref{eq:fsicen})
\begin{equation}
  \frac{\delta E_\mathrm{SIC}[\{\psi_\alpha\}]}
       {\delta U_\mathrm{SIC}^\mathrm{(local)}(\mathbf{r})}
  =
  0
  \quad.
\label{eq:OEPvar}
\end{equation}
Note that no additional ortho-normality constraint is needed in the
variation because it is already guaranteed by solving Eq.~(\ref{eq:OEP10}).

The variation is performed using the chain rule for functional
derivatives
$
{\delta E_\mathrm{SIC}}/{\delta U_\mathrm{SIC}^\mathrm{(local)}}
=
\left({\delta E_\mathrm{SIC}}/{\delta\varphi_i^*}\right)
\;
\left({\delta\varphi_i^*}/{\delta U_\mathrm{SIC}^\mathrm{(local)}}\right)
$
where the first factor represents the usual SIC mean-field and the
second factor the wave function response to varied local potential.
The detailed derivation is given in \cite{Mes08b}.  We obtain as a
final result an integral equation for the
$U_\mathrm{SIC}^\mathrm{(local)}$
\begin{subequations}
\label{eq:SIC-OEP-0}
\begin{equation}
  \sum_i \int \textrm d\mathbf{r'} \Big( 
     U_\mathrm{SIC}^\mathrm{(local)}(\mathbf{r'})-v^*_i(\mathbf{r'}) 
     \Big) 
    G_i(\mathbf{r},\mathbf{r'})\varphi_i^*(\mathbf{r'})
    \varphi_i(\mathbf{r}) 
    =
    0
\label{eq:statOEP10}
\end{equation}
where
\begin{equation}
   G_i(\mathbf{r},\mathbf{r'})
   =
   \sum_{j=1}^{+\infty}(1-\delta_{ij})
   \frac{\varphi_j^*(\mathbf{r})\varphi_j(\mathbf{r'})}{\epsilon_j-\epsilon_i}
\label{eq:G_i}
\end{equation}
is the single-particle Green function in the mean-field (\ref{eq:OEP10}).
The driving quantity in the integral equation~(\ref{eq:statOEP10}) is
the SIC potential with respect to the $\varphi_i$, namely
\begin{eqnarray}
  v_i(\mathbf{r})
  &=& 
  -\frac{1}{\varphi_i (\mathbf{r})} 
   \frac{\delta E_\mathrm{SIC}[\{\psi_\alpha\}] }
        {\delta \varphi_i^* (\mathbf{r})} 
  + \frac{1}{\varphi_i (\mathbf{r})}(\mathbf{r}|\hat{h}_\mathrm{LDA}|\varphi_i)
\nonumber\\
  &=& 
  \frac{1}{\varphi_i(\mathbf{r})}\sum_{\alpha} 
   u_{i \alpha}^{ *}U_\mathrm{LDA}[|\psi_\alpha|^2](\mathbf{r}) 
   \psi_\alpha(\mathbf{r})
   \quad=\quad
   \frac{1}{\varphi_i(\mathbf{r})}(\mathbf{r}|\hat{U}_\mathrm{SIC}|\varphi_i)
   \quad,
\label{eq:v_i-GS}
\end{eqnarray}
\end{subequations}
where the $\psi_\alpha$ are deduced from the $\varphi_i$ by the
unitary transformation (\ref{eq:unitrans}) whose coefficients
are determined by the symmetry condition (\ref{eq:symcond2}).
Note that we considered in this variation the diagonal basis states
$\varphi_i$ and the coefficients $\hat{u}$ of the unitary
transformation to the $\psi_\alpha$ as independent,
i.e. $\delta\hat{u}/\delta\varphi_i^*(\mathbf{r})=0$ (see also
section 4.4 of Ref. \cite{mes09} where we have shown that the
$\varphi_i$ and the $u_{i \alpha}$ should be considered as
independent in the variation of the energy or the action).

Eq.~(\ref{eq:statOEP10}) defines
$U_\mathrm{SIC}^\mathrm{(local)}(\mathbf{r})$ in a rather involved
way.  Its solution can be disentangled as
\begin{subequations}
\label{eq:SIC-OEP-OS}
\begin{eqnarray}
  U_\mathrm{SIC}^\mathrm{(local)} 
  &=&
  V_{\rm S}+V_{\rm K}+V_{\rm C}
  \quad,
\label{eq:OEP_gen10}
\\
  V_{\rm S} 
  &=& 
  \sum_i \frac{|\varphi_i|^2}{\rho}v_i
  \quad,
\label{eq:pot_slat10} \\
  V_{\rm K} 
  &=& 
  \sum_i \frac{|\varphi_i|^2}{\rho}
  (\varphi_i|U_\mathrm{SIC}^\mathrm{(local)}-v_i|\varphi_i)
  \quad,
\label{eq:pot_kli10} \\
  V_{\rm C} 
  &=& 
  \frac{1}{2}\sum_i \frac{\mathbf{\nabla}\cdot
  (p_i \mathbf{\nabla}|\varphi_i|^2)}{\rho}
  \quad,
\label{eq:pot_OEP10} \\
  p_i(\mathbf{r})
  &=&
  \frac{1}{\varphi_i^*(\mathbf{r})} \int \textrm d\mathbf{r'}
  \Big(U_\mathrm{SIC}^\mathrm{(local)}(\mathbf{r'})-v_i^*(\mathbf{r'})\Big) 
  \varphi_i^*(\mathbf{r'})G_i(\mathbf{r},\mathbf{r'})
  \quad.
\label{eq:p_i10}
\end{eqnarray}
\end{subequations}
From a practical point of view, this form is not simpler to use than
the original form (\ref{eq:statOEP10}) because the $V_{\rm K}$ and
$V_{\rm C}$ terms depend on the solution
$U_\mathrm{SIC}^\mathrm{(local)}$. However the separated
representation serves as a starting point to develop further approximations.

Some straightforward manipulations with the unitary transformation
(\ref{eq:unitrans}) allow to rewrite these quantities in terms of the
localized wave functions $\psi_\alpha$ as
\begin{subequations}
\label{eq:SIC-OEP-DS}
\begin{eqnarray}
  V_{\rm S} 
  &=& 
  \sum_\alpha \frac{|\psi_\alpha|^2}{\rho}
  U_\mathrm{LDA}[|\psi_\alpha|^2]
  \quad,
\nonumber\\
  V_{\rm K} 
  &=& 
  \frac{1}{\rho}\sum_{\alpha,\beta} 
  \Big( \sum_i |\varphi_i|^2 u_{i\alpha}^{*}u_{i\beta}
  \Big) 
  ({ \psi_\beta}|U_\mathrm{SIC}^\mathrm{(local)}
                -U_\mathrm{LDA} [|\psi_\alpha|^2]|\psi_\alpha)
  \quad,
\label{eq:pot_OEP3}\\
  p_i(\mathbf{r})
  &=&
  \frac{1}{\varphi_i^*(\mathbf{r})} \sum_\alpha\!
  u_{i\alpha}\int\!\!\textrm d\mathbf{r'} 
  \Big(U_\mathrm{SIC}^\mathrm{(local)}(\mathbf{r'})
       -U_\mathrm{LDA} [|\psi_\alpha|^2](\mathbf{r'})\Big) 
  \psi_\alpha^*(\mathbf{r'})G_i(\mathbf{r},\mathbf{r'})
  \quad.
\nonumber\\
\label{eq:pot_OEP2}
\end{eqnarray}
\end{subequations}
This expresses SIC-OEP in terms of the double-set representation
what we call the ``Generalized SIC-OEP'' formalism. Thus far it is
fully equivalent to the original SIC-OEP equations (\ref{eq:SIC-OEP-0})
and as involved to solve. But the double-set form
(\ref{eq:SIC-OEP-DS}) employs the more localized states $\psi_\alpha$
which produces a more forgiving hierarchy of importance for the
different terms.

It is to be noted that a very similar development is found in
\cite{Kor08}, but without addressing the feature of spatial
localization of the $\psi_\alpha$ when introducing the KLI
approximation.

\subsubsection{Generalized SIC-Slater approximation}
\label{sec:GSlat}

In this section, we show that the spatial localization of
the $\psi_\alpha$ permits to justify a powerful approximation.  We
define
\begin{eqnarray}
  F^\mathrm{(GS)}_\alpha 
  = 
  \Big(U_\mathrm{LDA}[|\psi_\alpha|^2]
  -\sum_\beta\frac{|\psi_\beta|^2}{\rho}U_\mathrm{LDA}[|\psi_\beta|^2]\Big)
  \psi_\alpha 
  \quad.
\label{eq:F_GS}
\end{eqnarray}
In the expected case that the $\psi_\alpha$ remain spatially
localized, we can assume that at each space point $\mathbf{r}$ 
one $\psi_\alpha$ dominates. This means that
\begin{eqnarray}
  F^\mathrm{(GS)}_\alpha
  \approx 
  0
  \quad. 
\label{eq:F_GS=0}
\end{eqnarray}
We now take up the ``Generalized SIC-OEP'' equations
(\ref{eq:SIC-OEP-DS}) and reshuffle them to display the
$F^\mathrm{(GS)}_\alpha$ explicitly.  The $V_{\rm S}$ term is dominating
compared to the $V_{\rm K}$ and $V_{\rm C}$ terms, although those
latter terms may be not a priori negligible.  Thus we assume
approximately
\begin{eqnarray}
  U_\mathrm{SIC}^\mathrm{(local)}
  \approx
  V_{\rm S} = \sum_\alpha \frac{|\psi_\alpha|^2}{\rho}U_\mathrm{LDA}[|\psi_\alpha|^2]
  \quad.
\label{eq:h_oep}
\end{eqnarray}
Inserting Eq. (\ref{eq:h_oep}) into Eqs. (\ref{eq:pot_OEP3}) and
(\ref{eq:pot_OEP2}), we obtain for the two remaining pieces
\begin{eqnarray}
  V_{\rm K}(\mathbf{r}) 
  &=& 
  -\frac{1}{\rho(\mathbf{r})} \sum_{\alpha,\beta} 
  \Big(\sum_i|\varphi_i(\mathbf{r})|^2u_{i\alpha}^{*}u_{i\beta}\Big)
  \int \textrm d\mathbf{r}' F^\mathrm{(GS)}_\alpha(\mathbf{r}')
  \psi_\beta^*(\mathbf{r}') 
  \quad,
\nonumber\\
  p_i(\mathbf{r})
  &=&
  - \frac{1}{\varphi_i^*(\mathbf{r})} 
  \sum_\alpha\!u_{i\alpha}\int\!\!\textrm d\mathbf{r'} 
  F^{(G)S*}_\alpha(\mathbf{r'}) G_i(\mathbf{r},\mathbf{r'})
  \quad.
\label{eq:vk_pi_gs}
\end{eqnarray}
Using the feature (\ref{eq:F_GS=0}) which follows from the
localization of the $\psi_\alpha$, we obtain 
\begin{eqnarray}
  && V_{\rm K}\approx 0 \quad,\quad
  p_i \approx 0 \Rightarrow V_{\rm C} \approx 0
  \quad.
\label{eq:pi=0}
\end{eqnarray}
This justifies a posteriori the assumption (\ref{eq:h_oep}) and so
allows to neglect the more involved contributions $V_{\rm K}$ and $V_{\rm C}$
\cite{Mes08b}.  Note that the double-set technique leaves full freedom
for the diagonal orbitals $\varphi_i$, whose degree of localization
can strongly vary according to the studied system (the $\varphi_i$
are, e.g., strongly delocalized in a metal and more localized in
covalent binding).

After all, the generalized SIC Slater (GS) approximation to SIC-OEP
\cite{Mes08b} can be summarized in the three coupled equations
\begin{subequations}
\begin{eqnarray}
   \Big(\hat{h}_\mathrm{LDA}-\hat{U}_\mathrm{GS}\Big)|\varphi_i) 
   &=&
   \epsilon_i |\varphi_i)
  \quad,
\label{eq:oep}
\\
  \hat{U}_\mathrm{GS} 
  &=& 
  \sum_\alpha \frac{|\psi_\alpha|^2}{\rho} \hat{U}_\mathrm{LDA}[|\psi_\alpha|^2]
  \quad,
\label{eq:h_GS}
\\
  0
  &=&
  (\psi_\alpha|\hat{U}_\mathrm{LDA}[|\psi_\alpha|^2]
   -\hat{U}_\mathrm{LDA}[|\psi_\beta|^2]|\psi_\beta)
  \quad.
\label{eq:symcond-GS}
\end{eqnarray}
\end{subequations}
Eq. (\ref{eq:oep}) determines the $\varphi_i$ for given
$\hat{U}_\mathrm{GS}$. Eq. (\ref{eq:symcond-GS}) determines the localized
states $\psi_\alpha$ by finding an appropriate unitary transformation
(\ref{eq:unitrans}). These $\psi_\alpha$ are employed in Eq.
(\ref{eq:h_GS}) to determine the local mean-field $\hat{U}_\mathrm{GS}$.
The coupled equations can be solved by iteration \cite{Mes08b}.

\subsubsection{Comment on the traditional SIC-Slater and SIC-KLI approximation}

One can show that the traditional SIC-OEP formalism
\cite{Kri92a,Kri92b} is obtained by the same reasoning as
previously, i.e. imposing that the diagonal orbitals $\varphi_i$
satisfy a local Schr\"odinger-like equation (\ref{eq:OEP10}).
But having no second set of wave functions at hand, it stops
at the stage (\ref{eq:SIC-OEP-OS}).
The traditional SIC-OEP equations \cite{Kri92a,Kri92b} are then
obtained replacing (\ref{eq:v_i-GS}) by
\begin{eqnarray}
v_i&=& U_\mathrm{LDA}[|\varphi_i|^2]
\label{eq:vi_oep_basic}
\end{eqnarray}
in the equations (\ref{eq:statOEP10}) and (\ref{eq:pot_slat10})-(\ref{eq:p_i10}).
The traditional SIC-Slater approximation is here formulated as
\begin{eqnarray}
U_\mathrm{S} &=& \sum_i \frac{|\varphi_i|^2}{\rho}U_\mathrm{LDA}[|\varphi_i|^2]
\label{eq:vi_sic_slat}
\end{eqnarray}
We discuss its validity in terms of the quantity
\begin{eqnarray}
  F^\mathrm{(S)}_i 
  = 
  \Big(U_\mathrm{LDA}[|\varphi_i|^2]
   -\sum_j\frac{|\varphi_j|^2}{\rho}U_\mathrm{LDA}[|\varphi_j|^2]\Big)
  \varphi_i
  \quad.
\label{eq:F_S}
\end{eqnarray}
One can shows that the Slater approximation 
is justified only if
\begin{eqnarray}
  F^\mathrm{(S)}_i\approx 0
  \quad, 
\label{eq:F_S=0}
\end{eqnarray}
i.e. only if the $\varphi_i$ orbitals remain spatially localized or
very delocalized (close to a Fermi gas). Similar reasoning applies to
the traditional KLI approximation.
But as extensively discussed previously, the $\varphi_i$ have no
particular reason to remain localized or to delocalized in the general
case. There are many favorable situations as, e.g., a tendency to
localized orbitals in organic molecules but also  strong delocalization in
metallic systems. Thus there is quite a choice of systems where the
Slater approximation is found to be applicable. However, there is also
a great number of systems where the Slater or KLI approximations fail.
This has been numerically shown in \cite{Kor08} for the example of
hydrogen chains.

The GS approximation contains with the double-set technique an extra
localization step which significantly enhances the range of validity
of the Slater approximation.  This was shown in terms of
several numerical examples in \cite{Mes08b,mes09-3} where, e.g., the
demanding hydrogen chain was found to be reasonably well
described within the GS approximation.  A key issue for
justifying a Slater-type approximation is that the single electron LDA
term $U_{\rm LDA}[\psi_\alpha]$ is close enough to the density
weighted average ($\rho_\alpha/\rho$) thereof.  The approximation
obviously works when the $\psi_\alpha$ are sufficiently localized,
as, for then, around the given point where $\psi_\alpha$ is
localized, one has $\rho \simeq \rho_{\alpha}$.  These strongly
localized orbitals correspond to a hydrogen or rare gas bond.  There
is the other extreme of metallic behavior in which all
$\psi_\alpha$ extend over the whole system and whose densities
resemble each other. This also provides an a priori well working
Slater approximation. In between these two extremes range numerous
conceivable cases. In particular, they can involve covalent
binding which are thus not so well approximated in a simple minded
Slater picture. We shall see below that the double set formulation
allows to address also such intermediate bindings and performs
well in these cases as well.

\subsection{Time-dependent formalism}

\subsubsection{Time-dependent Generalized SIC-OEP}

We now develop the time-dependent SIC-OEP and
``Generalized SIC-Slater'' formalisms \cite{Mes09-4}.  Starting
point is again the SIC quantum action (\ref{eq:varprinconstr}).
%, formulated in term of the $\psi_\alpha$ orbitals which satisfy the symetry
%condition (\ref{eq:td-symcond2}) (this will be implicit).  
We impose that the orbitals $\varphi_i$ satisfy a time-dependent
Schr\"odinger-like equation with local mean-field potential
$U_\mathrm{SIC}^\mathrm{(local)}(\mathbf{r},t) $, i.e.
\begin{equation}
  \Big( h_\mathrm{LDA}(\mathbf{r},t)
       -U_\mathrm{SIC}^\mathrm{(local)}(\mathbf{r},t) 
       - i \hbar \partial_t \Big) 
   \varphi_i(\mathbf{r},t) 
   =  0
   \quad.
\label{eq:Schrod}
\end{equation}
The optimal $U_\mathrm{SIC}^\mathrm{(local)}(\mathbf{r},t)$ is to be
determined by variation of the action 
\begin{equation}
  \frac{\delta A_{\rm SIC}}{\delta
    U_\mathrm{SIC}^\mathrm{(local)}(\mathbf{r},t)}
   =
   0
\label{eq:OEPvarAct}
\end{equation}
while the single-particle wave functions $\varphi_i$ become potential-dependent
quantities. No additional ortho-normalization constraint is needed in
this variation because it is already provided by the solution of
Eq. (\ref{eq:Schrod}).
The localized set $\psi_\alpha$ is again deduced from the $\varphi_i$
by the unitary transformation (\ref{eq:unitrans}) and the
transformation coefficients are to be determined by variation of the
action.  This yields once more the symmetry condition
(\ref{eq:symcond2}) to be fulfilled at each instant.  It is to
be noted that the emerging double set of $\varphi_i$ with
$\psi_\alpha$ is not exactly the same as the solution of TDSIC.
Nonetheless we use the same notations for sake of simplicity.

Similarly as in section \ref{sec:OEPstatderv}, the variation
(\ref{eq:OEPvarAct}) is again evaluated with the chain rule for
functional derivatives.  After a series of formal manipulations, one
obtains an integral equation for the optimal local mean-field
potential $U_\mathrm{SIC}^\mathrm{(local)}$
\begin{subequations}
\begin{eqnarray}
  &&
  0
  =
  \sum_i\!\int_{-\infty}^{t_1}\! \textrm dt'\!\int\!d\mathbf{r'} 
  \Big( U_\mathrm{SIC}^\mathrm{(local)}(\mathbf{r'},t')
        -v_i^*(\mathbf{r'},t') \Big) 
  K_i(\mathbf{r},t ; \mathbf{r'},t')
  \varphi_i^*(\mathbf{r'},t') \varphi_i(\mathbf{r},t) 
\nonumber\\
  &&
  \qquad
  +\mathrm{c.c.} 
  \quad,
\label{eq:tdOEP2}
\\
  &&
  K_i(\mathbf{r},t ; \mathbf{r'},t')
  =
  -i \sum_{j=1,j\ne i}^{+\infty}\varphi_j^*(\mathbf{r},t)
  \varphi_j(\mathbf{r'},t')\theta(t-t')
  \quad,
\label{eq:tdGF}
\\
  &&
  v_i(\mathbf{r},t)
  =
  -\frac{1}{\varphi_i(\mathbf{r},t)} 
    \frac{\delta}{\delta\varphi_i^* (\mathbf{r},t)}
    \int_{-\infty}^t \textrm dt' E_\mathrm{SIC}(t')  
   + 
  \frac{1}{\varphi_i(\mathbf{r},t)}(\mathbf{r}|\hat{h}_\mathrm{LDA}(t)|\varphi_i(t))
\nonumber\\
   &&
   \qquad\qquad
   =
   \frac{1}{\varphi_i(\mathbf{r},t)}\sum_{\alpha}u_{i \alpha}^{*}(t) 
   U_\mathrm{LDA} [|\psi_\alpha|^2](\mathbf{r},t) \psi_\alpha(\mathbf{r},t)
   \quad. 
\label{eq:v_i10_td}
\end{eqnarray}
\end{subequations}
As in the static case, we can decompose
$U_\mathrm{SIC}^\mathrm{(local)}$ in terms of separate contributions
\begin{subequations}
\begin{eqnarray}
  U_\mathrm{SIC}^\mathrm{(local)} 
  = 
  V_{\rm S} + \Re e \{ V_{\rm K}+V_{\rm C} \} - \Im{m} \{ V_{\rm TD1}+V_{\rm TD2} \}
\label{eq:TD-OEP_gen102}
\end{eqnarray}
where $V_{\rm S}$, $V_{\rm K}$, $V_{\rm C}$ are expressed exactly as in
(\ref{eq:pot_slat10})-(\ref{eq:pot_OEP10}) but where now the 
time dependence induces possible complex components we shall analyze
further below.  
However Eq. (\ref{eq:p_i10}) defining the $p_i$ for Eq. (\ref{eq:pot_OEP10})
is to be replaced by
\begin{eqnarray}
  p_i(\mathbf{r},t)
  &=& 
  \frac{\displaystyle
  \int_{-\infty}^t \!\!\textrm dt' \int \!\textrm d\mathbf{r'} 
       \big(U_\mathrm{SIC}^\mathrm{(local)}(\mathbf{r'},t')
            \!-\!v_i^*(\mathbf{r'},t')\big)
           \varphi_i^*(\mathbf{r'},t') K_i(\mathbf{r},t ; \mathbf{r'},t')
   }{\varphi_i^*(\mathbf{r},t)}
\label{eq:p_i-td}
\end{eqnarray}
The potentials $V_{\rm S}$, $V_{\rm K}$, $V_{\rm C}$, which also appear in the stationary case,
are now complemented by  two dynamical contributions
\begin{eqnarray}
  V_{\rm TD1} 
  &=& 
  \frac{1}{\rho}\sum_i\frac{\mathbf{\nabla}^2|\varphi_i|^2}{4} 
  \int_{-\infty}^t \textrm dt'(\varphi_i(t')|v_i(t')|\varphi_i(t')) \ ,
\label{eq:pot_oep_td10}
\\
  V_{\rm TD2} 
  &=& 
  \frac{1}{\rho}
  \sum_i\Big(|\varphi_i|^2\frac{\partial p_i}{\partial t} 
         +\mathbf{J}_i.\mathbf{\nabla} p_i   \Big) \ ,
\label{eq:pot_oep_td10b}
\end{eqnarray}
\end{subequations}
where 
$\mathbf{J}_i=
\frac{1}{2i}(\varphi_i^*\nabla\varphi_i-\varphi_i\nabla\varphi_i^*)$ 
is the current density.  Note that the potential $V_{\rm TD1}$ contains a
time integral, thus memory effects, while the potential $V_{\rm TD2}$
involves the time derivative of the $p_i$.

The standard way to derive the time-dependent Slater and
time-dependent KLI approximations starts from the above separation in
terms of the propagating basis $\varphi_i$.  More robust
approximations will be obtained from a formulation in terms of the
localizing set $\psi_\alpha$. The separation can be remapped
to
\begin{subequations}
\begin{eqnarray}
  && 
  V_{\rm S} 
  = 
  \sum_\alpha\frac{|\psi_\alpha|^2}{\rho}U_\mathrm{LDA}[|\psi_\alpha|^2]
  \quad,
\\
  && 
  V_{\rm K} 
  = 
  \frac{1}{\rho}\sum_{\alpha,\beta} 
  \Big(\sum_i |\varphi_i|^2
  u_{i\alpha}^{*}u_{i\beta}\Big) 
  ({ \psi_\beta}|U_\mathrm{SIC}^\mathrm{(local)}
                -U_\mathrm{LDA} [|\psi_\alpha|^2]|\psi_\alpha) \ ,
\label{eq:pot_OEP2-td0} \\
  &&
  p_i(\mathbf{r},t)
  = 
  \frac{1}{\varphi_i^*(\mathbf{r},t)}\sum_\alpha u_{i\alpha}(t)
  \int_{-\infty}^t \textrm dt'\int \textrm d\mathbf{r'} 
\nonumber\\
  &&
  \hspace*{4.5em}
  \Big(U_\mathrm{SIC}^\mathrm{(local)}(\mathbf{r'},t')
       -U_\mathrm{LDA} [|\psi_\alpha|^2](\mathbf{r'},t')\Big) 
  \psi_\alpha^*(\mathbf{r'},t') K_i(\mathbf{r},t;\mathbf{r'},t')
\nonumber\\
\label{eq:pot_OEP2-td}
\\
  &&
  (\varphi_i|v_i|\varphi_i)
  =
  \sum_{\alpha,\beta}u_{i\beta}u_{i\alpha}^*
  (\psi_\beta|U_\mathrm{LDA}[|\psi_\alpha|^2]|\psi_\alpha) \ ,
\nonumber\\
  &&
  \hspace*{4.5em}
  =
  \sum_{\alpha,\beta}u_{i\alpha}u_{i\beta}^*
  (\psi_\alpha|U_\mathrm{LDA}[|\psi_\alpha|^2]|\psi_\beta)
  \quad=(\varphi_i|v_i|\varphi_i)^* \ .
\label{eq:vtd1}
\end{eqnarray}
\end{subequations}
The last equation (\ref{eq:vtd1}) implies 
$\Im{m}\{(\varphi_i|v_i|\varphi_i)\}=0$
which, in turn, yields $\Im{m}\{V_{\rm TD1}\}=0$. This thus allows to
remove this term in the decomposition (\ref{eq:TD-OEP_gen102}).

These equations for the optimal local mean-field together with the
time-dependent mean-field equation (\ref{eq:Schrod}) and with the
symmetry condition (\ref{eq:symcond2}) constitute TDSIC-OEP in double
set representation. Its solution is by no means simpler than the
solution of fully fledged TDSIC. But the equations with explicit
separation of the optimal local mean-field provide a good starting
point for approximations.

\subsubsection{Time-dependent generalized SIC-Slater approximation}
\label{sec:TDGSderiv}

The reasoning to derive a time-dependent generalized Slater
approximation proceeds very similar to the static case (section
\ref{sec:GSlat}). We introduce the function
$F^\mathrm{(GS)}_\alpha(\mathbf{r},t)$ defined in (\ref{eq:F_GS}) and assume the
generalized Slater approximation
\begin{eqnarray}
  U_\mathrm{SIC}^\mathrm{(local)}
  \approx
  V_{\rm S} 
  =
  \sum_\alpha\frac{|\psi_\alpha|^2}{\rho}U_\mathrm{LDA}[|\psi_\alpha|^2]
  \quad.
\label{eq:h_oep-td}
\end{eqnarray}
Inserting into Eqs. (\ref{eq:pot_OEP2-td0}) and
(\ref{eq:pot_OEP2-td}) yields
\begin{eqnarray*}
  && 
  V_{\rm K} 
  = 
  -\frac{1}{\rho}\sum_{\alpha,\beta} 
  \Big(\sum_i |\varphi_i|^2
  u_{i\alpha}^{*}u_{i\beta}\Big) 
  \int \textrm d\mathbf{r}
  F^\mathrm{(GS)}_\alpha(\mathbf{r},t)\psi_\beta^*(\mathbf{r},t) 
  \quad\approx 0\ ,
\\
  && 
  p_i(\mathbf{r},t)
  = 
  -\frac{1}{\varphi_i^*(\mathbf{r},t)}\sum_\alpha u_{i\alpha}(t) 
  \int_{-\infty}^t \textrm dt' \int \textrm d\mathbf{r'} 
  F^\mathrm{(GS)*}_\alpha(\mathbf{r'},t') K_i(\mathbf{r},t ; \mathbf{r'},t')
  \approx 0 \ ,
\\
  && 
  \Rightarrow\quad V_\mathrm{TD2} \approx 0
  \quad,\quad V_\mathrm{C}\approx 0 \ .
\label{eq:pi=0_td}
\end{eqnarray*}
These results are consistent with the approximation (\ref{eq:h_oep-td}).
Thus the potential (\ref{eq:h_oep-td}) is probably a good
approximation of the time-dependent SIC-OEP potential for a broad
class of problems, as for example when the localization by the symmetry
condition (\ref{eq:symcond2}) works well.  The time-dependent
generalized Slater potential (\ref{eq:h_oep-td}) has the same form as
the stationary GS potential. It does not contain memory
effects anymore, which is another consequence of the localization of
the $\psi_\alpha$.

The emerging scheme is called time-dependent generalized SIC Slater
approximation (TDGS). It can be summarized by the coupled equations
\begin{subequations}
\begin{eqnarray}
  \Big(\hat{h}_\mathrm{LDA} - {U}_\mathrm{GS}\Big)|\varphi_i) 
  &=&
  \mathrm{i}\hbar\partial_t|\varphi_i)
  \quad,
\label{eq:oep-td}
\\
  {U}_\mathrm{GS} 
  &=&
   \sum_\alpha \frac{|\psi_\alpha|^2}{\rho}U_\mathrm{LDA}[|\psi_\alpha|^2]
  \quad,
\label{eq:h_GS_td}
\\
  \forall t\;:\quad
  0
  &=&
  (\psi_\alpha|U_\mathrm{LDA}[|\psi_\alpha|^2]
              -U_\mathrm{LDA}[|\psi_\beta|^2]|\psi_\beta)
  \quad,
\label{eq:symcondTDGS}
\end{eqnarray}
\end{subequations}
where the $\psi_\alpha$ are obtained from the $\varphi_i$ by the
unitary transformation (\ref{eq:unitrans}) which makes the
$\psi_\alpha$ to satisfy the symmetry condition
(\ref{eq:symcondTDGS}).

For the same reasons as discussed in the static case, TDGS should
represent an improvement to conventional time-dependent SIC-Slater and
SIC-KLI approximations to the extent that it opens a larger class of
problems for which the approximation is applicable. This will be
demonstrated on practical test cases in  section \ref{sec:numres}.

\subsubsection{Conservation law I: Energy conservation}
\label{sec:conserv_GS-E}

Within TDGS, the $\varphi_i$ orbitals propagate under the influence of the
potential (\ref{eq:h_GS_td}) according to Eq. (\ref{eq:oep-td}).
The total energy is computed with $E_\mathrm{SIC}$ as given in Eq.
(\ref{eq:fsicen}). We remind that variation of  $E_\mathrm{SIC}$
defines
the SIC mean-field $\hat{U}_{\rm SIC}$ as defined in (\ref{eq:Usic})~:
\begin{eqnarray}
  \frac{\delta }{\delta \varphi_i^*(\mathbf{r},t')} 
  E_\mathrm{SIC}[\{\psi_\alpha\}](t) 
  &=& 
  (\mathbf{r}|\hat{h}_{\rm SIC}(t')|\varphi_i(t'))\;\delta(t-t')
  \quad,
\nonumber\\
  \hat{h}_{\rm SIC}(t)
  &=& 
  \hat{h}_\mathrm{LDA}(t) - \hat{U}_{\rm SIC}(t)
  \quad.
\end{eqnarray}
Energy conservation is an issue for time-independent external fields,
i.e. for $\partial_t v_{ext}=0$. The time evolution of the energy thus
becomes
\begin{eqnarray}
  \partial_t E_\mathrm{SIC}
  &=&
  \sum_i \int \textrm dt' \int \textrm d \mathbf r 
  \partial_t\varphi_i^*({\bf r},t')
  \frac{\delta }{\delta\varphi_i^*({\bf r},t')}E_{\rm SIC}[\{|\psi_\alpha|^2\}](t)
  +
  \mbox{c.c.}
\nonumber\\
  &=&
  \sum_i \int \textrm dt' 
  \big(\partial_t\varphi_i(t')\big|
  \hat{h}_{\rm SIC}(t')\varphi_i(t')\big)\;\delta(t-t')
  +
  \mbox{c.c.}
\nonumber\\
  &=&
  \frac{\mathrm{i}}{\hbar} \sum_i\Big[
  \big(\hat{h}_{\rm GS}(t)\varphi_i(t)\big|\hat{h}_{\rm SIC}(t)\varphi_i(t)\big)
  -
  \big(\hat{h}_{\rm SIC}(t)\varphi_i(t)\big|\hat{h}_{\rm GS}(t)\varphi_i(t)\big)
  \Big]
\nonumber\\
  &=&
  \frac{2}{\hbar}\Im{m} \Big\{
  \sum_i\big(\frac{\mathbf{p}^2}{2m}\varphi_i\big|\hat{U}_{\rm SIC}\varphi_i\big)
  -\sum_i\big(\frac{\mathbf{p}^2}{2m}\varphi_i\big|\hat{U}_{\rm GS}\varphi_i\big)
  \Big\}
\nonumber\\
  &=&
  \frac{2}{\hbar}\Im{m}\Big\{
  -\frac{\hbar^2}{2m}\int \textrm d\mathbf{r} 
  \sum_\alpha U_\mathrm{LDA}[|\psi_\alpha|^2] \psi_\alpha\Delta\psi_\alpha^*
\nonumber\\
  &&\qquad\qquad
  +\frac{\hbar^2}{2m}\int \textrm d\mathbf{r}\sum_\beta \frac{|\psi_\beta|^2}{\rho}U_\mathrm{LDA}
  [|\psi_\beta|^2]\sum_i \varphi_i\Delta\varphi_i^*
  \Big\}
\nonumber\\
  &=&
  \frac{\hbar}{m}\Im{m} \Big\{
  -\int \textrm d\mathbf{r} \sum_\alpha 
  U_\mathrm{LDA}[|\psi_\alpha|^2] \psi_\alpha\Delta\psi_\alpha^*
\nonumber\\
  &&\qquad\qquad
  +\int \textrm d\mathbf{r}\sum_\beta \frac{|\psi_\beta|^2}{\rho}U_\mathrm{LDA}
  [|\psi_\beta|^2]\sum_\alpha \psi_\alpha\Delta\psi_\alpha^*
  \Big\}
\nonumber
\end{eqnarray}
We finally obtain the time variation of the SIC energy
(\ref{eq:fsicen}) within a GS propagation
\begin{eqnarray}
  \partial_t E_\mathrm{SIC}[\{\psi_\alpha\}]
  &=&
  - 
  \frac{\hbar}{m}\Im{m}\Big\{\sum_\alpha\int \textrm d\mathbf{r} 
   F^\mathrm{(GS)}_\alpha(\mathbf{r},t)\Delta\psi_\alpha^*(\mathbf{r},t)
  \Big\}
\label{eq:cons_en_GS}
\end{eqnarray}
where we employed the deviation function $F^\mathrm{(GS)}_\alpha$ from
Eq. (\ref{eq:F_GS}). The relation (\ref{eq:cons_en_GS}) shows that the
energy is not strictly conserved. The quality of energy conservation
depends on the quality of the generalized Slater approximation because
the deviation is driven by the same function $F^\mathrm{(GS)}_\alpha$
which enters the decision on negligible terms at the end of section
\ref{sec:TDGSderiv}.

We compare this result with the traditional time-dependent SIC-Slater
propagation.  We can show similarly that the time variation of the
associated total energy $E_\mathrm{SIC}[\{\varphi_i\}]$, now expressed
in terms of the diagonal orbitals $\varphi_i$ reads
\begin{eqnarray}
  \partial_t E_\mathrm{SIC}[\{\varphi_i\}]
  =
  - \frac{\hbar}{m}\Im{m}\Big\{
     \sum_i \int \textrm
     d\mathbf{r}F^\mathrm{(S)}_i(\mathbf{r},t)\Delta
     \varphi_i^*(\mathbf{r},t) \Big\}
\end{eqnarray}
with $F^\mathrm{(S)}_i$ given by Eq. (\ref{eq:F_S}).  Energy
conservation holds for the traditional SIC-Slater scheme if
$F^\mathrm{(S)}_i\approx 0$, thus if the physical system as a whole
remains homogeneous or very localized.  
The extra localization for the
$\psi_\alpha$ in the double-set technique makes it very likely that
energy conservation is improved for TDGS. It is known for traditional
TDKLI and TD-Slater that energy conservation lasts for a certain time
interval after which energy explodes \cite{Mun07a}. We expect that
TDGS has a similar behavior but with a much extended time interval of
practical energy conservation, which allows to use it in a wider range 
of physical situations. 

\subsubsection{Conservation law II: Zero Force Theorem}
\label{sec:conserv_GS-ZFT}

The Zero Force Theorem (ZFT) states that a time variation of the
total electron momentum can be caused only by an "external" potential
\cite{Lev85a,Vig95a,VBar05,Mun07a}, i.e.
\begin{equation*}
  \partial_t \langle \mathbf{P} \rangle 
  = 
  - \int{d}\mathbf{r}\rho\nabla v_{\rm ext}
  \quad.
\end{equation*}
It stems from the fact that the electron-electron interaction is
translational invariant and can not produce a "net" force on the
system which, in turn, leads to the ZFT in the form \cite{Vig95a}
\begin{eqnarray}
  \forall \tilde\rho \quad : \qquad 
  \int{d}\mathbf{r}\tilde\rho(\mathbf{r},t) 
  \nabla U_{\rm mf}[\tilde\rho](\mathbf{r},t) 
  = 
  0
\label{eq:ZFT-gen3}
\end{eqnarray}
where $U_{\rm mf}$ is the local mean-field potential of the considered
method. The ZFT holds for the LDA mean-field and the ADSIC one.
We now are going to check the ZFT for the TDGS mean-field
$U_\mathrm{GS}$.

The time evolution of the total momentum is given by
\begin{eqnarray}
  \partial_t \sum_i(\varphi_i|\mathbf{p}|\varphi_i)
  &=&
  \sum_i \int \mathrm d \mathbf{r} 
  \Big\{
  (\mathbf{r}|\hat{v}_{\rm ext} + \hat{U}_\mathrm{LDA} 
   - \hat{U}_\mathrm{GS}|\varphi_i)^*\nabla\varphi_i(\mathbf{r})
\nonumber\\
  &&\qquad\qquad
   + 
  (\mathbf{r}|\hat{v}_{\rm ext} + \hat{U}_\mathrm{LDA} 
  - \hat{U}_\mathrm{GS}|\varphi_i)\nabla\varphi_i^*(\mathbf{r})
  \Big\}
\nonumber
\\
  &=& 
  \int{\mathrm d}\mathbf{r}  v_{\rm ext} \nabla\rho 
  + 
  \int{\mathrm d}\mathbf{r} U_\mathrm{LDA}[\rho]\nabla\rho 
\nonumber\\
  && 
  - 
  \int{\mathrm d}\mathbf{r} \sum_\alpha 
  \frac{|\psi_\alpha|^2}{\rho} U_{\rm LDA}[|\psi_\alpha|^2] \nabla\rho
\label{eq:ZFT-gs1}
\end{eqnarray}
The second term disappears as can be shown by a partial integration
and exploiting the ZFT for $U_\mathrm{LDA}$.
We add 
$0=\sum_\alpha 
\int{\mathrm d}\mathbf{r}U_\mathrm{LDA}[|\psi_\alpha|^2]\nabla|\psi_\alpha|^2$
to the third term and reshuffle Eq. (\ref{eq:ZFT-gs1}) to
\begin{eqnarray}
  \partial_t \sum_i(\varphi_i|\mathbf{p}|\varphi_i) 
  &=& 
  -\int{\mathrm d}\mathbf{r}\rho\nabla v_{\rm ext} 
  -2\Re{e}\Big\{
     \sum_\alpha\int{\mathrm d}\mathbf{r} 
     F^\mathrm{(GS)}_\alpha(\mathbf{r},t)\nabla\psi_\alpha^*(\mathbf{r},t)
   \Big\}
\label{eq:viol_zft_td-GS}
\end{eqnarray}
where we employ $F^\mathrm{(GS)}_\alpha$ from Eq. (\ref{eq:F_GS}).
Again, we see that the deviation function $F^\mathrm{(GS)}_\alpha$
drives also the term that violates the ZFT. The ZFT is well fulfilled if
$F^\mathrm{(GS)}_\alpha$ is small, i.e. if TDGS is valid. In reverse,
violation of ZFT and energy conservation is a valuable indicator for
the breakdown of TDGS.

With similar steps we can evaluate the ZFT for the traditional
time-dependent Slater approximation and obtain
\begin{eqnarray}
  \partial_t \sum_i(\varphi_i|\mathbf{p}|\varphi_i) 
  &=& 
  - \int \textrm d\mathbf{r} \rho \nabla v_{\rm ext} 
  - 2 \Re{e}\Big\{ 
       \sum_i\int{\mathrm d}\mathbf{r} 
       F^\mathrm{(S)}_i(\mathbf{r},t) \nabla \varphi_i^*(\mathbf{r},t)
      \Big\}
\end{eqnarray}
where we use the $F^\mathrm{(S)}_i$ from Eq. (\ref{eq:F_S}).  The ZFT is
thus verified within a traditional SIC-Slater propagation only if
$F^\mathrm{S}_i\approx 0$, thus if the physical system remains
homogeneous or very localized.
We have argued above that the range of $F^\mathrm{(S)}_i\approx 0$ is
much smaller than the range of $F^\mathrm{(GS)}_\alpha\approx 0$.
This means that TDGS should maintain the ZFT for a longer time span
than traditional time-dependent SIC-Slater.

\subsection{Alternative localization criteria}
\label{sec:loccrit}

One major effect of the symmetry condition (\ref{eq:symcondTDGS}) is
that it produces states $\psi_\alpha$ which are better localized than
the originally given $\varphi_i$. This was the particular feature
which we employed to motivate TDGS. On the other hand, the symmetry
condition is very expensive to solve in practical calculations. Thus
it is worth trying to achieve better localization by less demanding
equations. There exist many localization criteria \cite{Fos60,Edm63}.
After a series of numerical tests with many  of these localization
criteria, we have found as a best compromise for a localization
criterion the spatial variances of the one-body orbitals~:
\begin{equation}
  \overline{\Delta r^2}_\mathrm{sp}
  = 
  \sum_\alpha\left[
  (\psi_\alpha|\mathbf{r}^2|\psi_\alpha)
  -
  (\psi_\alpha|\mathbf{r}|\psi_\alpha)^2
  \right]
  \quad,
\label{eq:variance}
\end{equation}
where the index ``sp'' stands for the summed single-particle variances.
Minimization of this variance yields the localization equations
\begin{equation}
  0
  =
  (\psi_\alpha|\overline{\mathbf{r}_\alpha}
              -\overline{\mathbf{r}_\beta}|\psi_\beta)
  \quad,\quad
  \overline{\mathbf{r}_\alpha}
  =
  (\psi_\alpha|\mathbf{r}|\psi_\alpha)
\label{eq:loccond}
\end{equation}
which then replaces the symmetry condition (\ref{eq:symcondTDGS}) in
the TDGS equation. It serves to determine the coefficients of the
unitary transformation (\ref{eq:unitrans}). It is again a non-linear
equation which has to be solved iteratively. But the expectation value
$\overline{\mathbf{r}_\alpha}$ can be computed much faster than the
Coulomb field. Thus TDGS with the localization condition
(\ref{eq:loccond}) is computationally less demanding. We have to see
how it performs in practice.

\section{Numerical results}
\label{sec:numres}

\subsection{Brief reminder of the various studied formalisms}

In the following, we will compare the results obtained with (TD)GS and
other approaches to those obtained with full (TD)SIC as a benchmark. The
corresponding mean-field Hamiltonians are summarized in table
\ref{tab:schemes},
all being used in one-body Schr\"odinger-like equations of the form 
$\hat{h}|\varphi_i)=\mathrm{i}\hbar\partial_t|\varphi_i)$.
\begin{table}[htbp]
\begin{center}
\begin{tabular}{|l|l|c|}
\hline
\rule[-3pt]{0pt}{16pt}
expression of $\hat{h}$ in $\hat{h}|\varphi_i)=\epsilon_i|\varphi_i)$&
method & acronym\\ 
\hline
\hline
\rule[-5pt]{0pt}{20pt}
$\hat{h}_{\rm LDA}[\rho]$ & LDA & LDA\\
\hline
\rule[-13pt]{0pt}{32pt}
$\displaystyle
\hat{h}_{\rm LDA}[\rho]-\hat{U}_{\rm LDA} \left[ \frac{\rho}{N}
  \right]$ & Average Density SIC & ADSIC\\ 
\hline
\rule[-18pt]{0pt}{40pt}
$\displaystyle
\hat{h}_{\rm LDA}[\rho]-\sum_j \frac{|\varphi_j|^2}{\rho} \hat{U}_{\rm
  LDA}\left[|\varphi_j|^2\right]$ & Standard SIC Slater & Slat\\ 
\hline
\rule[-18pt]{0pt}{40pt}
$\displaystyle
\hat{h}_{\rm LDA}[\rho] - \sum_\alpha \frac{|\psi_\alpha|^2}{\rho}
\hat{U}_{\rm LDA}\left[|\psi_\alpha|^2\right]$ & Generalized SIC
  Slater & \begin{minipage}{5em}GS(sym)\\GS(var)\end{minipage}
\\
$\displaystyle
0
  =
  (\psi_\beta|U_{\rm LDA} [ |\psi_\beta|^2 ]-U_{\rm LDA} [|\psi_\alpha|^2 ]|\psi_\alpha)
$
& & \\
\hline
\rule[-17pt]{0pt}{32pt}
$\displaystyle
 \hat{h}_{\rm LDA}[\rho] - \sum_\alpha \hat{U}_{\rm
  LDA}\left[|\psi_\alpha|^2\right] |\psi_\alpha)(\psi_\alpha|$ &
  full SIC (benchmark) &
  (TD)SIC \\ 
$\displaystyle
0
  =
  (\psi_\beta|U_{\rm LDA} [ |\psi_\beta|^2 ]-U_{\rm LDA} [|\psi_\alpha|^2 ]|\psi_\alpha)
$
& & \\
\hline
\end{tabular}
\end{center}
\vspace*{6pt}
\caption{\label{tab:schemes}
The hierarchy of mean-field Hamiltonians, from simple-most LDA (top
line) to full TDSIC (bottom line). The right column shows the acronyms
used in the figures and discussion.
}
\end{table}
Note that for GS the symmetry condition (\ref{eq:symcond2}) should be added
for the two last schemes, to define the localized states $\psi_\alpha$
required in the corresponding Hamiltonians. 
As an alternative, we consider
the localization criterion (\ref{eq:loccond}) derived from
minimization of the spatial variance (\ref{eq:variance}), see
section \ref{sec:loccrit}. We abbreviate the scheme using the symmetry
condition as ``GS(sym)'' and the alternative scheme using Eq.
(\ref{eq:loccond}) as ``GS(var)''.

The static and dynamical calculations are performed on 3D
coordinate-space grid using standard techniques, for details see
e.g. \cite{Cal00,Rei03a}. The calculations are  restricted to valence
electrons. They are the $3s$ electrons in Na, the $2s$ and $2p$
electrons in C, and naturally the $1s$ electron in H. The coupling of
the ionic cores to the valence electrons is described by
pseudopotentials.  For the C and H atoms, we use Goedecker-type
pseudopotentials \cite{Goe96} and for Na atoms the soft local
pseudopotentials of \cite{Kue99}.  The LDA part employs the
exchange-correlation energy functional from \cite{Per92}. ADSIC is
performed as explained in \cite{Leg02}.  The static solution is done
by accelerated gradient iteration \cite{Rei82,Blu92}.  Time stepping
is done by fourth order Taylor expansion of the exponential
propagation operator \cite{Dav85}.  The Poisson equation is solved by
a fast Fourier technique combined with separate treatment of the
long-range terms~\cite{Lau94}. Polarizabilities are computed from two
static calculations where one is performed under the influence of
a small static external dipole field.

\subsection{Potential energy surfaces}

The C$_2$ molecule is found to be a critical test case. The electronic
structure changes substantially from the spin-saturated, covalently
bound dimer ground state to the highly spin polarized asymptotic
atomic states. It is demanding for a theory to describe this transition
smoothly. 
\begin{figure}[htbp]
\begin{center}
\includegraphics[width=0.7\linewidth]{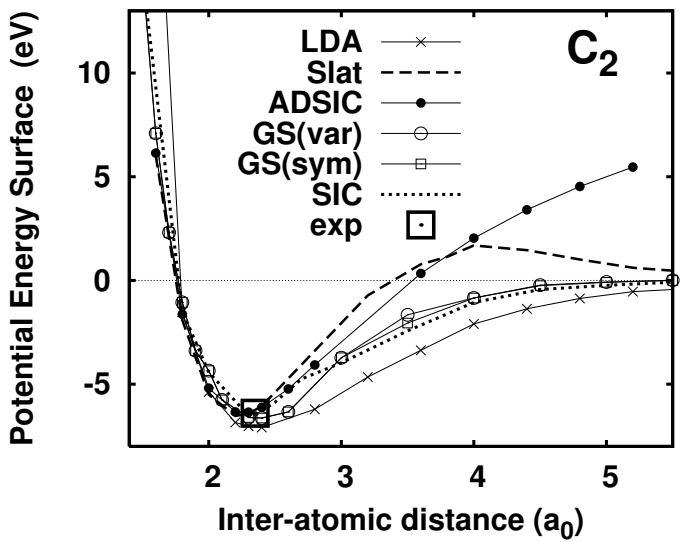}
\caption{Potential energy surface of C$_2$ for various SIC
  calculations as indicated.
%rloc=0.3
\label{fig:C2}}
\end{center}
\end{figure}
Figure \ref{fig:C2} shows the Born-Oppenheimer potential energy surface for
the C$_2$ dimer computed with a variety of approaches. Let us first start with
the LDA approach which provides a qualitatively good approach with a fair
reproduction of both bond length and dissociation energy but which
unfortunately {underestimates the ground state vibration frequency by about
  25~\%.}  All SIC corrected methods provide a much better reproduction of the
bond length and the dissociation energy of the equilibrium state (less than
5~\% of discrepancy with respect to the experimental data). They also improve
the value of the vibration frequency~: Slater and ADSIC yield an agreement
within less than 5~\%, and full SIC, GS(var) and GS(sym) within typically
10--15~\%. Note also that GS(var) deliver results which are almost identical
to GS(sym) while being much less expensive numerically. This will also hold
in the dynamical regime (see Sec.~\ref{sec:dyn}).
  
However one observes strange behaviors in
ADSIC and Slater at intermediate distance with the appearance of a
totally unphysical "bump" in the potential energy surface. The
effect is not present neither in full SIC nor in GS approximations
which thus both provide a correct account of the potential energy
surface. The defect observed in Slater and ADSIC has different
origin. In Slater, it is probably to be attributed to a conflict
between a tendency of the system to create "delocalized" orbitals, to
ensure bonding, and a tendency towards "localized" orbitals, to ensure
a better account of the SIC. The two set strategy proves here very
valuable by resolving the conflict. The ADSIC problem comes from the
fact that asymptotically the ADSIC correction should take a form
different from that at smaller distance because of the different
number of involved electrons (4 in each separate C atom, 8 in the
dimer).  ADSIC requires compact systems and is generally not suited
for describing fragmentation.

\subsection{Polarizabilities}

Polarizabilities are a sensitive test case for density functional
approaches
\cite{Cha95b,Gua95,Gis98,Kue00,Faa02,Koe07,Pem08,mes09-3}. We will
thus discuss this issue here for three sufficiently different systems.
\begin{figure}[ht]
\begin{center}
\includegraphics[width=\linewidth]{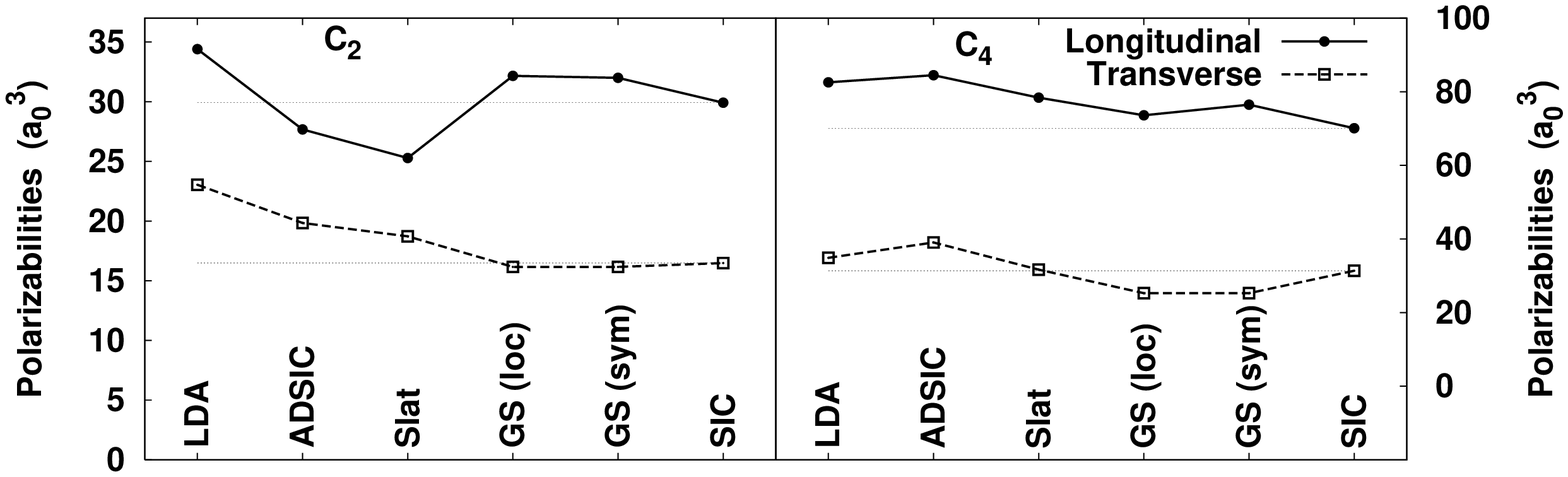}
\caption{
Transverse and longitudinal polarizabilities of the C$_2$ molecule
(left) and the C$_4$ chain (right), calculated in various SIC
schemes. Horizontal lines emphasize the SIC benchmark values and ease
the comparison with the other results.
\label{fig:polC}}
\end{center}
\end{figure}
As a first example, we consider the carbon molecules C$_2$ and C$_4$.
Figure \ref{fig:polC} shows their polarizabilities for the various
approximations. To put the subsequent results on C molecules into
perspective, we recall the computed polarizations for the C atom:
along $z$ axis, $\alpha_z=10.40$ ${a_0}^3$ for both SIC and GS,
along $x,y$ axes, $\alpha_{x,y}= 11.52$ ${a_0}^3$ for GS and 11.76
${a_0}^3$ for SIC. Experimental values for the molecular
polarizabilities seem not be available. But one can compare with other
computed values obtained with much different methods
\cite{Bia02,Abd02} They yield generally comparable values.  In
\cite{Bia02}, the longitudinal polarizability for C$_2$ is
$\alpha_\parallel=$ 25 ${a_0}^3$ for the ab initio methods and 34
${a_0}^3$ for LDA/GGA, while the transverse one is $\alpha_\perp=$25
${a_0}^3$ or 100 ${a_0}^3$ respectively, the latter value being a
strange exception. The results for C$_4$ are $\alpha_\parallel=$ 92 or
94 ${a_0}^3$ and $\alpha_\perp=$ 30 or 32 ${a_0}^3$. Our results are
generally lower for $\alpha_\perp$. However, it is to be noted that
our calculations differ in the employed functionals and
pseudopotentials which both can have a sensitive influence on the
results. In view of that, the comparison as a whole looks satisfying.

The main aim of figure \ref{fig:polC} is a comparison of methods
within the same setup. The C$_2$ dimer shows the larger
variance of the results and is obviously more critical than the
C$_4$ chain. It is obvious that GS provides the best approximations
to full SIC and it is interesting that both versions, GS(sym) and
GS(var), perform almost equally well.

\begin{figure}[ht]
\begin{center}
\includegraphics[width=\linewidth]{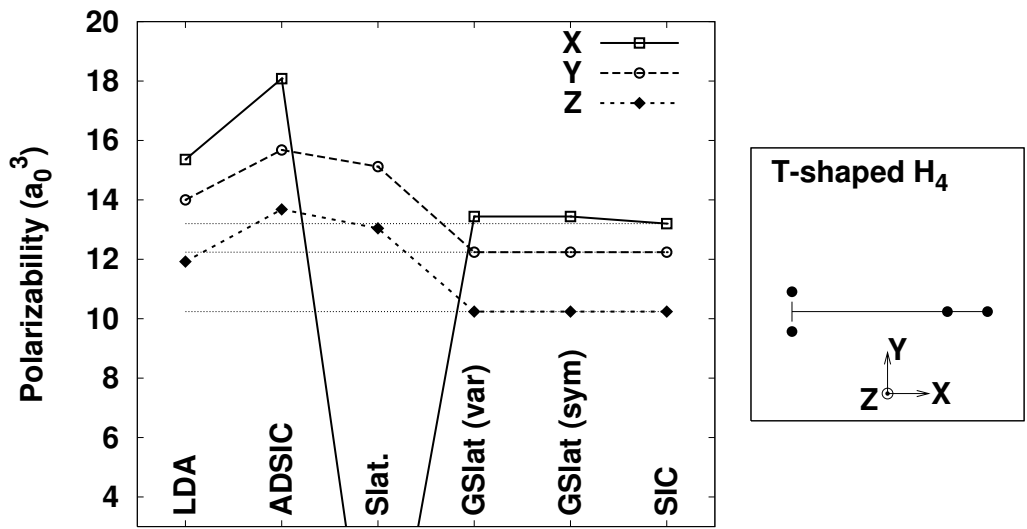}
\caption{
Polarizabilities, for various SIC schemes as indicated, of the ground
state configuration H$_4$ in the T-shaped configuration, displayed in
the right panel. Horizontal lines emphasize the SIC benchmark
values and ease the comparison with the other results.
\label{fig:polT-H4}}
\end{center}
\end{figure}
The H$_4$ ground state configuration is a ``T-shaped'' molecule
\cite{Die00} as indicated in the right panel of figure
\ref{fig:polT-H4}.  The H$_4$ ground state configuration consists of
two H$_2$ dimers bound with a H$_2$-H$_2$ center of mass distance of
6.425 $a_0$.  This is a demanding configuration as it contains two
well localized cloud of electrons at each H$_2$ center loosely
connected between the centers. Traditional SIC-Slater and KLI tends to
delocalize the wave functions too much.  The triaxial spatial
configuration provides three different polarizabilities depending on
the orientation of the external electric field relative to the
molecule.  The left panel of figure \ref{fig:polT-H4} compares the
results for the three polarizabilities.  Again we see that both
variants of GS come very close to the benchmark (SIC), while LDA,
as well as ADSIC, overestimate the polarizabilities, and traditional
SIC-Slater is totally off. The overestimation is related to an
exaggerated delocalization for LDA and ADSIC. The failure of the
traditional Slater approximation indicates a too strong localization
of the bonds. GS finds the right compromise.

\begin{figure}[ht]
\begin{center}
\includegraphics[width=\linewidth]{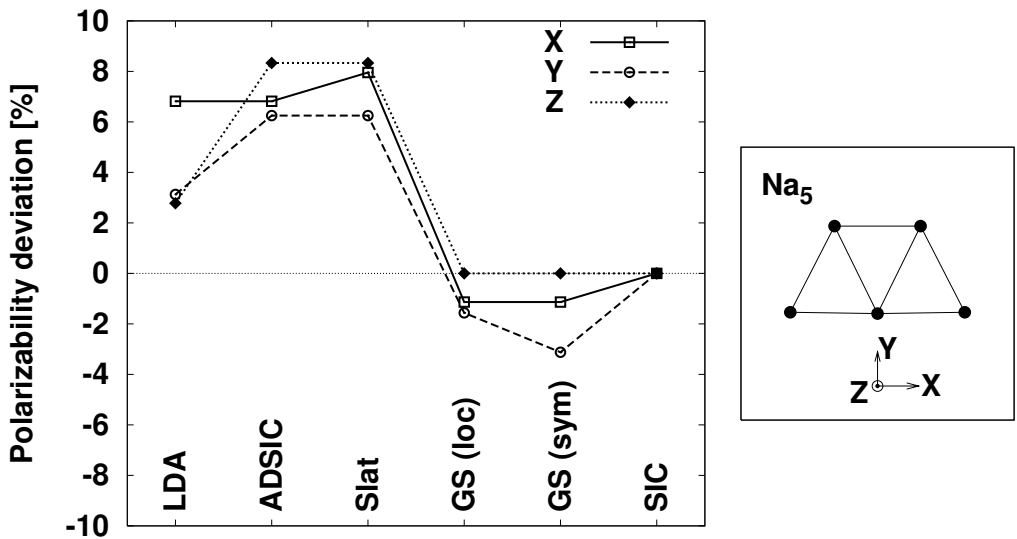}
\caption{
Polarizabilities, for various SIC schemes as indicated, of Na$_5$,
displayed in the right panel. Horizontal lines emphasize the SIC
benchmark values and ease the comparison with the other results.
\label{fig:na5pol}}
\end{center}
\end{figure}
As a final test case for polarizability, we consider the small sodium
cluster Na$_5$ representative of simple metallic systems. Being a piece
of a metal this cluster should have a delocalized electron cloud. But
the rather soft binding of Na$_5$ degrades metalicity and drives
towards weak localization. This makes Na$_5$ a particularly critical
test case amongst metallic clusters \cite{Mun07a}.  The cluster is
planar (see the right panel of figure \ref{fig:na5pol}) which
corresponds to a triaxial shape and leads to three rather different
polarizabilities along the three major axes of the system. In order 
to display in a better readable way the various results, we have thus 
chosen to present values relative to the full SIC ones rather than absolute 
values.  The left
panel of figure \ref{fig:na5pol} shows the
polarizabilities. We obtain much larger absolute values of
polarizabilities than in the case of organic systems due to the
metallic nature of bonding (delocalization and lower binding). Not
surprisingly, all approaches perform rather well, better than in
organic systems, as is to be expected for a simple metallic system.
Within the lower error bands, we still see differences in the
performance with a clear improvement provided by both GS versions.

\subsection{Time-dependent case}
\label{sec:dyn}

As discussed in sections \ref{sec:conserv_GS-E} and \ref{sec:conserv_GS-ZFT}
Generalized Slater does not exactly fulfill conservation laws and has thus to 
be used with caution in actual time dependent processes. Still it is 
interesting to test in dynamical scenarios, especially in the linear domain 
where it could advantageously replace more complicated approaches. 
The term linear domain refers to electronic oscillations with small
amplitudes. In 
the context of clusters and molecules,  it largely refers to 
the analysis of the optical response which plays  a key 
role in a broad variety of dynamical scenarios, both in the linear and
non linear  
domains \cite{Rei03a}. It thus represents a key issue in these systems. 
TDDFT, in particular in its real time formulation \cite{Cal97b}, 
is especially well suited to address such phenomena. For then, the
point is simply  
to excite the system with a sufficiently small energy, whatever its
value, so that dynamics sets in and  
allows a spectral analysis \cite{Cal97b}. We shall thus mostly discuss
such cases in the following.

\subsubsection{Na$_5$}
As a first test case for dynamics, we consider Na$_5$ which was found to
be critical probe for studying conservation properties \cite{Mun07a}
because of its soft and easily polarizable electron cloud.  
We excite the electronic cloud by applying a boost in the $x$
direction to each wave function. This simulates a very short laser
pulse. We compare the case of very small excitation in the linear regime
with that of a larger excitation. No absorbing boundary is used here,
so that the total energy should be conserved in time.

\begin{figure}[htbp]
\begin{center}
\includegraphics[width=\linewidth]{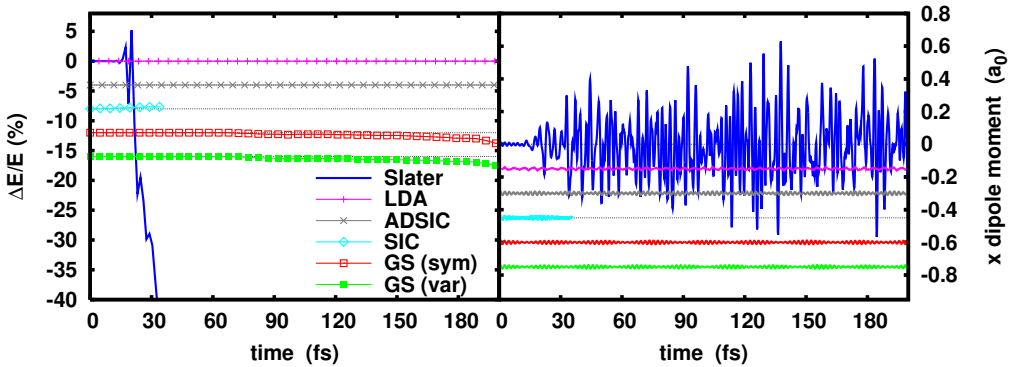}
\caption{
(Color online)
Time evolution of the total energy (left) and the dipole
moment in the spatial  $x$  direction (right) for Na$_5$ {after a
boost  of the wave functions with
momentum  0.001/a$_0$ in the $x$ direction}, for various SIC
calculations as indicated. For the sake of clarity, some results have
been down-shifted by a constant offset.
\label{fig:NAen001}}
\end{center}
\end{figure}

We first start with the low excitation case presented in
Figure~\ref{fig:NAen001}.
The left panel compares energy conservation. In variational
approaches, the total energy is conserved. We thus plot the deviation
to this energy conservation, that is $\Delta E/E = \left[E(t)
  - E(0)\right]/E(0)$.
As expected, 
LDA and TDSIC show up as straight lines because
these methods are proven to conserve energy.
TDGS will finally also develop an energy
instability but it stays stable for much longer more than 7
times the standard Slater approximation which diverges after
only 30 fs. Moreover, we see that the faster variant
GS(var) performs as well as the version employing the involved
symmetry condition. The right panels of figure \ref{fig:NAen001} show
the time evolution of dipole moments. For the sake of clarity, the
signals for GS(sym) and GS(var) have been shifted. In agreement
with the time evolution of the total energy, in a TDGS calculation,
the dipoles do not exhibit any significant evolution, while a standard
Slater calculation produces large oscillations.

\begin{figure}[htbp]
\begin{center}
\includegraphics[width=0.7\linewidth]{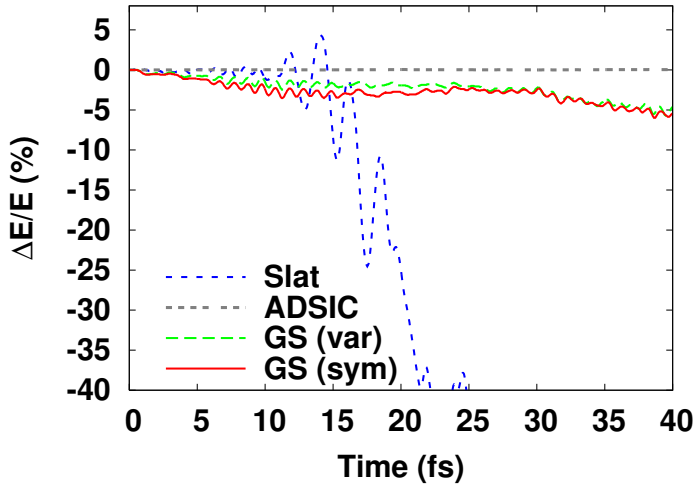}
\caption{
(Color online)
Total energy deviation of Na$_5$ {for a boost momentum of 0.05/a$_0$
 in the
  $x$ direction.}
%Na5 ; Boost=0.05 ; NoAbsorbingBounds ; total energy.
\label{fig:NAen05}}
\end{center}
\end{figure}

Figure \ref{fig:NAen05} shows the energy conservation for the same
test case Na$_5$ but for a much larger excitation energy
%$E^*=0.17$~eV in the non-linear regime. 
{50 times higher, thus in the non-linear regime.}
We see again that both variants of TDGS provide a longer stability
time than the standard Slater approximation. However, the stability
time is much shorter than in the previous case of small
excitation. The quality of Slater approximations is degrading with
increasing excitation. The applicability has to be checked for each
system and dynamical  range anew.

\subsubsection{T-shape H$_4$}

\begin{figure}[htbp]
\begin{center}
\includegraphics[width=0.9\linewidth]{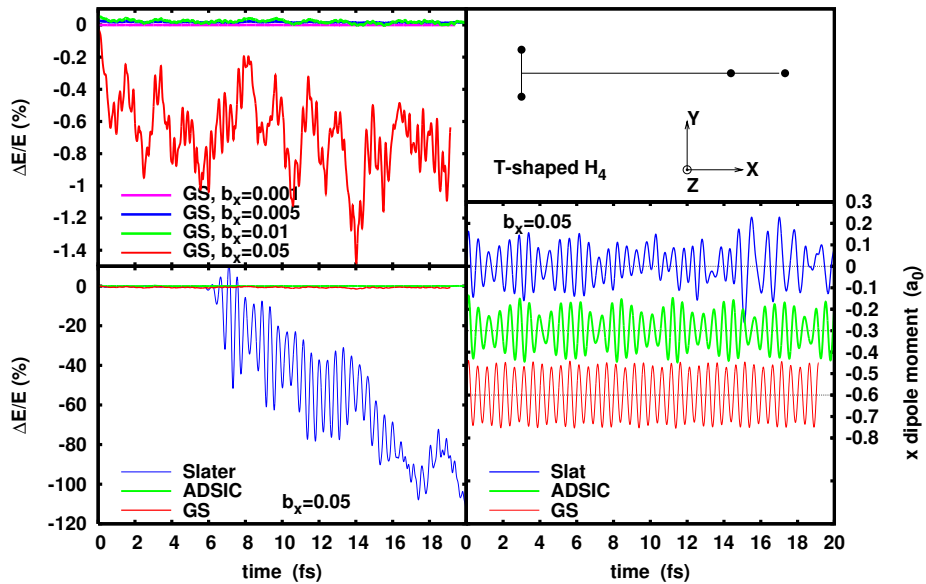}
\caption{
(Color online)
T-shaped H$_4$ (ionic configuration in the top right panel)
  excited by {boosts $b_x$ in the $x$ direction as indicated}. Left
  column: total energy deviation as a function of 
  time; bottom right panel: time evolution of the $x$ dipole moment.
\label{fig:h4-T}}
\end{center}
\end{figure}

We now turn to the case of H$_4$ in the $T$ configuration, as displayed
in the top right panel of Figure~\ref{fig:h4-T}. 
We excite this cluster the same way, that is with an 
instantaneous boost. Since GS(var) and GS(sym) turn out to give
very close results, we only plot one of these results and denote them
as GS without mentioning the criterion used for the calculation of the
unitary transform.
The findings from the lower two panels are the same as from the
previous results. The standard Slater approximation runs rather quickly
into violation of energy conservation, while GS stays stable for much
longer. The instability in energy shows up at later times in the
dipole signal. The upper left panel explores energy conservation for
GS at various initial excitation energies.
It is apparent that time span at which GS propagates in stable manner
depends crucially on the excitation energy of the process.
At the present stage of development it can be safely used only in the
linear domain.

\section{Conclusion}
\label{sec:concl}

We have presented in this paper a generalized formulation of
SIC-OEP in the time domain. It relies on the introduction of a
double set of electron orbitals. The double-set strategy plays a key
role for the time propagation of full SIC but also provides in the
OEP context a valuable tool for deriving approximate versions of the
full SIC-OEP, in stationary as well as time dependent processes. It
allows in particular to introduce the Generalized Slater (GS)
approximation which preserves the simplicity of the standard Slater
approximation and yet extends its range of validity. While a first
set of "in general localized" orbitals serves to fulfill
ortho-normality and to construct the approximate Hamiltonian, a
second set of "physical" wavefunctions allows a simple time
propagation. We have shown that the GS improves over the defects
of the simple Slater approximation in particular what concerns
conservation laws but still does not allow to fulfill them exactly. We
have thus performed various tests on a series of clusters and
molecules in order to explore in a practical way the capabilities
of GS. We have seen that it performs quite well for static
properties and allows to deal with dynamics in the linear domain.  It
then provides a simple and transparent alternative to the full
TD-OEP or TD-SIC which require substantial numerical effort. The
numerical performance of the method still require to be
optimized. We have shown that the expensive symmetry condition may be
advantageously replaced by the conceptually and computationally
simpler "localization" ansatz which opens up possibly new,
simplified, approaches to the problem. This strategy nevertheless
still requires to be explored in more detail. Work along that line
is in progress.

\bigskip

\noindent
Acknowledgment: This work was supported by a research scholarship from the
Alexander-von-Humboldt foundation.

\bibliographystyle{elsart-num}
\bibliography{biblio}

\end{document}